\documentclass[aps,prl,superscriptaddress,amsfonts,amsmath,amssymb,twocolumn,showpacs,floatfix]{revtex4}
\usepackage{url}
\usepackage{bm}
\usepackage{graphicx}
\usepackage{amsmath}
\usepackage{amstext}
\usepackage{amssymb}
\usepackage{amsfonts}
\usepackage{amsbsy}
\usepackage{verbatim}
\usepackage{color}
\usepackage[colorlinks=true, urlcolor=blue, linkcolor=blue, citecolor=blue, pdftex]{hyperref}
\usepackage{multirow}

\newcommand{\id}{{\mathbf 1}}

\include{diagrammatics}

\begin{document}

\title{Fractionalization of itinerant anyons in one dimensional chains}

\author{Didier Poilblanc}
\affiliation{Laboratoire de Physique Th\'eorique UMR-5152, CNRS and
Universit\'e de Toulouse, F-31062 France}

\author{Matthias Troyer}
\affiliation{Institut f\"ur Theoretische Physik, ETH-H\"onggerberg, 8093 Z\"urich, Switzerland}

\author{Eddy Ardonne}
\affiliation{Nordita, Roslagstullsbacken 23, SE-106 91 Stockholm, Sweden}

\author{Parsa Bonderson}
\affiliation{Station Q, Microsoft Research, Santa Barbara, CA 93106-6105, USA}

\date{\today}

\begin{abstract}
We construct models of interacting itinerant non-Abelian anyons moving along one-dimensional chains, focusing in particular on itinerant Ising anyon chains,
and derive effective anyonic $t$-$J$ models for the low energy sectors. Solving these models by exact diagonalization, we find a fractionalization of the anyons into charge and (non-Abelian) anyonic 
degrees of freedom --  a generalization of spin-charge separation of electrons which occurs in Luttinger liquids. A detailed description of the excitation spectrum by combining spectra for charge and anyonic sectors requires a subtle coupling between charge and anyonic excitations at the microscopic level (which we also find to be present in Luttinger liquids), despite the macroscopic fractionalization.
\end{abstract}

\maketitle

One of the most striking phenomena that can emerge as a result of confining electrons to a one-dimensional
system is the dissociation of the electrons' spin and charge into two
independent degrees of freedom. This effect, called ``spin--charge separation,'' was first predicted by
Anderson~\cite{anderson}. The theory describing this behavior of one-dimensional electrons was developed
by Tomonaga \cite{t50} and Luttinger \cite{l63}, and
Haldane later introduced the concept of Luttinger liquids \cite{h81}. 

For two dimensional systems, fractionalization of the electron can also occur in topologically ordered phases of matter~\cite{wen-91}, in particular fractional quantum Hall states. The emergent quasiparticle excitations of these strongly interacting electron liquids are anyons~\cite{Leinaas77}, and may possess fractional electric charge~\cite{Laughlin83} and exotic exchange statistics characterized by the braid 
group~\cite{arovas-84prl722,bonderson-2011aa}.
The anyonic properties of the quasiparticles may be encoded in topological quantum numbers, often called ``topological charges.'' 
Of particular interest are non-Abelian anyons~\cite{non-abelian-anyons}, which possess a collective topological Hilbert space that grows (roughly) exponentially with the number of quasiparticles and braiding operations, represented by matrices, that act upon this degenerate state space in a non-commuting fashion. Such non-Abelian anyons, which may occur 
in quantum Hall systems~\cite{moore-read,read-rezayi,Lee07,Levin07,Bonderson07d}, $p$-wave superconductors and solid state heterostructures~\cite{majoranas}, 
could provide topologically fault-tolerant platforms for quantum information processing~\cite{kitaev-2003}.

Even though the electrons' quantum numbers are fractionalized in the quasiparticles of topologically ordered systems, the (fractional) electric charge carried by a quasiparticle is directly correlated with its topological charge. Thus, a question that immediately comes to mind is: Do itinerant non-Abelian anyons in one dimension undergo a process analogous to spin-charge separation of itinerant electrons?
In this Letter, we will study this basic, but important question by introducing
simple  models of interacting itinerant non-Abelian anyons.
We conclusively show that mobile non-Abelian anyons in one dimension can indeed
exhibit a separation of their charge and anyonic degrees
of freedom analogous to spin-charge separation of electrons.


The most direct example of spin-charge separation for electrons was seen in the $t$-$J$ model~\cite{ZhangRice}, which can be obtained from the Hubbard model \cite{Hubbard_model}
in the limit of strong on-site repulsion $U/t\gg 1$, where $t$ is the nearest-neighbor (NN) hopping.
In this low-energy effective model, where the high-energy doubly occupied local states are integrated out,
there is an antiferromagnetic interaction $J=4t^2/U$ between two electrons on neighboring sites. To motivate our choice of models for studying itinerant anyons, we note that the (subtle) $J=0$ limit, corresponding to the $U=\infty$ Hubbard model, is already  of  interest: using the Bethe-Ansatz form of 
the ground state (GS)
wavefunction, it was shown~\cite{ogata-shiba,maekawa} that its charge degrees of freedom can be expressed as a Slater determinant of spinless fermions with antiperiodic boundary conditions (equivalent to hard-core bosons), while its spin degrees of freedom are equivalent to those of a one-dimensional S=1/2 Heisenberg model.

\paragraph{The model --}
To investigate whether similar phenomena occur in anyonic systems, we introduce models for itinerant anyons, and derive an effective $t$-$J$ model for the low energy excitations. While our construction is general, we specifically consider Ising-type anyons
appearing in the fractional quantum Hall candidate states likely to occur in the second 
Landau level~\cite{moore-read,Lee07,Levin07,Bonderson07d} (particularly at $\nu=5/2$),  $p+ip$-superconductors
and heterostructures~\cite{majoranas}.

To physically motivate our model, we will use the quantum Hall effect setting, in particular
the quasiparticle excitations of the $\nu=1/2$ Moore-Read state \cite{moore-read}. These excitations carry both electric and 
topological charges.
The electric charges occur in multiples of $e/4$, while the topological
charges (or anyon types) are those of the Ising theory, namely $(\id,\sigma,\psi)$
which follow the fusion rules
\begin{align*}
\sigma\times\sigma &=\id+\psi & \sigma \times \psi &=\sigma & \psi \times \psi = \id
\end{align*}
(fusion is symmetric, and fusion with the vacuum $\id$ is trivial, $\id \times x = x$, for any $x$).
The different types of quasiparticle excitations that can occur in the Moore-Read state follow from repeated
fusion of the ``fundamental'' quasihole $(\sigma,e/4)$ and/or quasielectron $(\sigma,-e/4)$,
leading to $(\id,\frac{n e}{2})$, $(\psi,\frac{n e}{2})$, and $(\sigma,\frac{(n+1/2) e}{2})$ for $n$ integer 
(where $(\psi,e)$ is the electron, while $(\id,2e)$ can be identified with the vacuum).
Of these quasiparticles, we will only consider those with the smallest electric charges 0 and $e/4$, as the other excitations will be penalized by large charging energies. 
Thus, the effective model can be described in terms of $(\psi,0)$ and $(\sigma,e/4)$ quasiparticle types only. With a suitable electric potential, we can arrange that the neutral $(\psi,0)$ particle is higher in energy than the charged $(\sigma,e/4)$ \cite{note1}, so we do not allow localized $(\psi,0)$ quasiparticles.

\begin{figure}[t!]
\vskip 0.3truecm
\includegraphics[width=1.0\columnwidth]{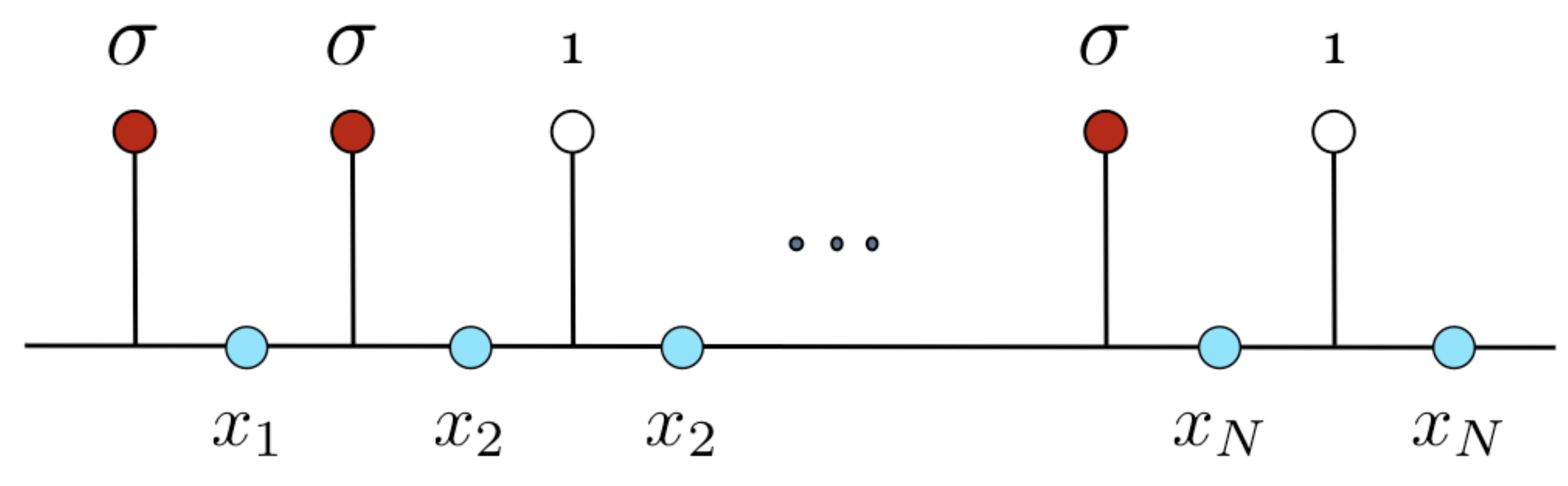}
\caption{A chain with $N$ itinerant anyonic quasiparticles. 
The (red) dark dots represent the sigma quasiparticles, open circles represent empty sites, and the (light) shaded 
dots describe the fusion channels of the non-local states and can take the value $\id$, $\sigma$ and $\psi$, as dictated 
by the fusion rules. }
\label{fig:chain}
\end{figure}

We consider a chain of $L$ sites, each of which can only be occupied by a ``hole,'' i.e. an empty site labeled by the ``vacuum''  $(\id,0$), or by a single $(\sigma,e/4)$ anyon. Neutral $(\psi,0)$ anyon types can tunnel
between the sites.  
Although the number of anyons is only conserved modulo 4 (four anyons can form an electron that can be added or removed from the condensate) 
we consider a fixed density $\rho$ of quasiparticles, since states with a different density will be at much higher energies.
In addition, we will drop the superfluous charge labels from now on.
In the golden chain model~\cite{ftl07} and its generalizations, which correspond to $\rho=1$,
the anyons are immobile, because of a hardcore constraint. For $\rho\ne 1$, basis states of the anyonic chain (see Fig.~\ref{fig:chain})
are all positions of the $N=\rho L$ anyons combined with all admissible labelings of the fusion tree with $x_i \in \{\id,\sigma,\psi\}$ that satisfy the constraints given by the fusion rules.
Throughout this letter, we will
assume periodic boundary conditions $x_L=x_0$ and no overall topological charge constraint, meaning the chain of anyons is in fact imbedded on a torus.

Hopping of $\sigma$ anyons is restricted to NN sites and given, for the $j$-th anyon, by
\begin{equation}
-t \;\; \Bigl| \raisebox{-5mm}{\includegraphics[height=1cm]{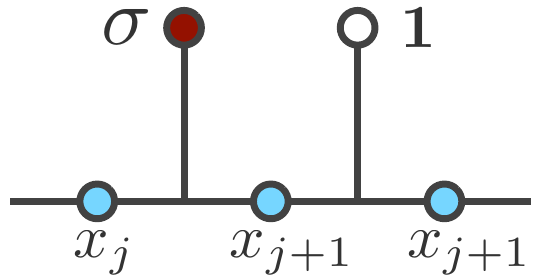}}\Bigr\rangle
\Bigl\langle \raisebox{-5mm}{\includegraphics[height=1cm]{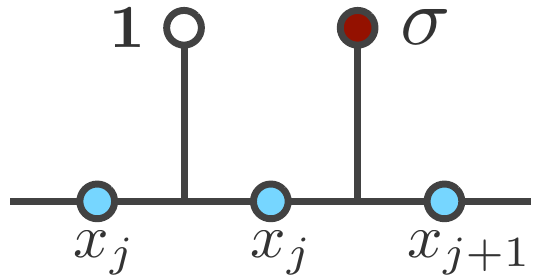}}\Bigr|
+ {\rm h.c.},
\label{eq:ht}
\end{equation}
where $x_j , x_{j+1}$ can take the values $\id$, $\sigma$, or $\psi$ (and the trivial action of this operator on 
the unshown fusion tree elements is left implicit). The Hamiltonian includes a sum over the entire fusion tree 
of such hopping operators.

When two (charged) $\sigma$ anyons occupy NN sites, they
experience a Coulomb repulsion $V$ 
and an exchange interaction
$J$, (as occurs in the golden chain model \cite{ftl07},) that favors the fusion channel of these two $\sigma$ anyons to be
the trivial particle $\id$ when $J>0$ and the fermion $\psi$ when $J<0$. The exchange interaction term originates from virtual processes, including all orders and quasiparticle types (even those we are neglecting for being at higher energies), that transfer topological charge between the localized anyons without changing their localized topological charge values~\cite{Bonderson09b}. (In the case of Ising-type anyons, the only possible non-trivial topological charge that can be transferred in this manner is the neutral $\psi$ charge.) This is similar to a Heisenberg exchange term arising from a Hubbard interaction when integrating out double occupied sites in the derivation of the fermionic $t$-$J$ model. These two interaction terms can be combined as
\begin{align}
&(V\delta_{z,z'}-J/2) \;\;
\Bigl| \raisebox{-5mm}{\includegraphics[height=1cm]{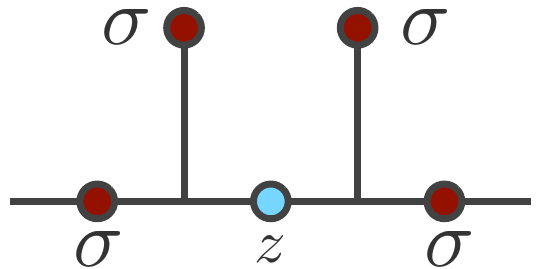}}\Bigr\rangle
\Bigl\langle \raisebox{-5mm}{\includegraphics[height=1cm]{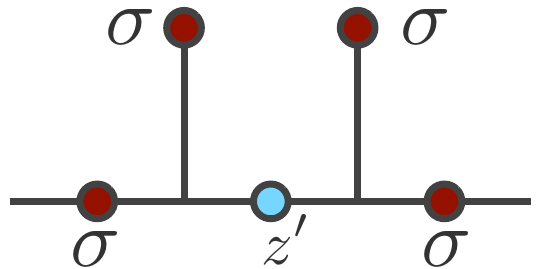}}\Bigr|
\nonumber
\\
&+(V-J\delta_{x,y}) \;\;
\Bigl| \raisebox{-5mm}{\includegraphics[height=1cm]{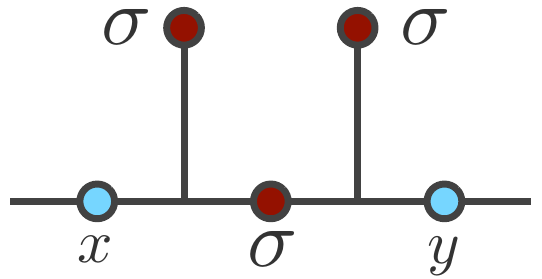}}\Bigr\rangle
\Bigl\langle \raisebox{-5mm}{\includegraphics[height=1cm]{xy}}\Bigr|
\label{eq:hJ}
\end{align}
where  $x,y,z,z'$ can take the values $\id$ or $\psi$. As in Eq.~(1), 
the Hamiltonian includes a sum 
over the entire fusion tree of these interaction terms. 
All other matrix elements are zero.
We refer to the model as the Ising anyon $t$-$J$ chain.

{\it Phases of the $\rho=1$ localized anyon chain --}
We first briefly recall the
behavior of the pure Ising anyon
chain, when there is exactly one $\sigma$ anyon localized on each site of the chain.
In this case, the local constraints impose
an exact alternation of frozen $\sigma$ bonds with fluctuating $\id$ or $\psi$ bonds.
For these later bond variables, the terms in Eq.~(\ref{eq:hJ}) are exactly those
of the transverse field Ising model at the critical point, described by the Ising
conformal field theory (CFT) with central charge $c=1/2$.~\cite{difrancesco}

\begin{figure}[t!]
\vskip 0.3truecm
\includegraphics[width=1.0\columnwidth]{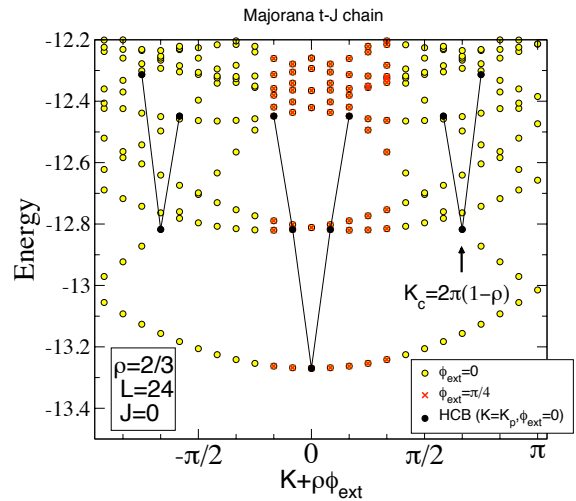}
\caption{Low-energy spectrum of a 24-site Ising anyon chain at $J=0$ and density $\rho=2/3$.
Shaded (yellow) circles correspond to the spectrum at $\phi_{\rm ext}=0$ as a function of
momentum $K$. The (red) crosses correspond to the spectrum at $\phi_{\rm ext}=\pi/4$ restricting
$K\in[-\pi/3,\pi/6]$ (so that ${\tilde K}\in[-\pi/6,\pi/3]$). The parent charge excitations are
shown by solid (black) circles.
}
\label{fig:Maj0}
\end{figure}

{\it Spectrum for $J=0$ --}
We next consider the simple case where $J=V=0$. Although the exchange interaction $J$ is absent, it is
important to keep in mind that the particles that hop along the chain are
anyons, whose state is described by the fusion tree labels $(x_1,..., x_N)$.
If we forget
about this labeling, our model with $J=0$ is equivalent to that of
itinerant hardcore bosons (HCB).
However, the labeling makes the anyons effectively distinguishable and introduces degeneracies.
On an open chain, one simply gets $2^{N/2}$ copies of the same HCB spectrum. In contrast, 
around a torus,
hopping an anyon across the ``boundary'' cyclically translates the labels of the fusion tree.
To recover the same labeling, in general, all $N$ particles have to
be translated over the boundary. Thus, one anyon
hopping over the boundary has the same effect as a phase shift $\phi_k=\frac{2\pi k}{N}$ (with $k$
an integer) or equivalently an Abelian U(1) flux.
The same effect occurs for electrons with spin, but so far has been discussed only in the special case of a
single hole in a spin background~\cite{parola-sorella-92}.
Hence, the complete $J=0$ anyonic spectrum (at zero
external flux)
is given by the union of the HCB spectra $E_{\rm charge}(p,\phi_k)$
taken at all discrete values of $\phi=\phi_k$ (with extra degeneracies as evidenced later on), where
one can use the mapping to spinless fermions (with an extra $\pi$-flux for $N$ even),
\begin{equation}
E_{\rm charge}(p,\phi)=-2t\sum_{j(p)}\cos{(\frac{2\pi}{L}(j+\frac{1}{2})+\frac{\phi}{L})}\, ,
\label{Eq:charge0}
\end{equation}
and $\{j(p)\}$ is a set of $N$ integers
characterizing the $\phi=0$ HCB eigenstates (hereafter named ``parent states" and labelled by $p$) of momenta
$K_p=\frac{2\pi}{L}\sum_{j(p)}(j+\frac{1}{2})$.
The states labelled by $p$ and $k$ carry momenta
$K=K_p+\rho\phi_k$.
From the above considerations, it is then clear that each parent state 
of the HCB spectrum is
extended into a discrete set of exponentially many degenerate levels on a parabola -- the same parabola that
one gets by adding flux. We have verified this numerically (see Fig.~\ref{fig:Maj0}) 
using Lanczos exact diagonalizations (ED).
Interestingly, the dispersion of the parent states, $E_{\rm charge}(p,0)$
vs. $K_p$, reveals linear ``charge'' modes at momenta $K=0$ and $K_c=2\pi(1-\rho)$.
Note that the $J=0$ spectrum does not depend on the nature of the particles
(e.g. electrons, Ising anyons, or even more general anyons), except in determining the degeneracies of the states.

When the anyons experience an external flux $\phi_{\rm ext}$,
it is added to the ``internal'' phase shift $\phi_k$,
i.e. $\phi=\phi_{\rm ext}+\phi_k$, and the energy spectrum consequently depends continuously on the ``pseudo-momentum"
${\tilde K}=K_p+\rho\phi=K+\rho\phi_{\rm ext}$ (see Fig.~\ref{fig:Maj0})~\cite{note2}.

\begin{figure}[t!]
\vskip 0.3truecm
\includegraphics[width=1.0\columnwidth]{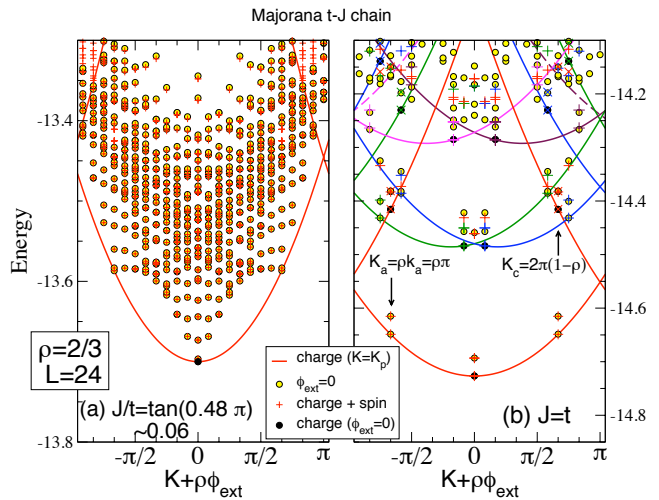}
\caption{Low-energy spectra of a 24-site Ising anyon $t$-$J$ chain at density $\rho=2/3$
(a) for small $J/t$ and (b) for $J=t$. The $\phi_{\rm ext}=0$ spectra are shown by shaded (yellow) circles
as a function of momentum $K$.
The parent charge excitations ($K=K_p$ and $\phi_{\rm ext}=0$) are shown by black dots.
The corresponding charge excitation branches (obtained by varying $\phi_{\rm ext}$) are shown by continuous lines
of different colors.
The $+$ and $\times$ symbols correspond to the 
sum of the charge and
(expected) anyonic excitation spectra (see text). The colors of these symbols are the same as their corresponding charge excitation parabola.
}
\label{fig:Maj}
\end{figure}

{\it Analysis of the anyon spectra at $J\ne 0$ --}
We now consider nonzero $J$ and examine the case $J>0$ and $V=0$, for simplicity.
Typical low-energy spectra of Ising anyon chains
obtained by Lanczos ED are shown in Figs.~\ref{fig:Maj}(a,b).
To understand such spectra, let us first focus on the low-energy region,
where a small finite $J$ lifts the degeneracies of the $J=0$ charge excitation parabola by 
an energy proportional to $JL$.
When $J\sim t/L^2$, the spectra originating from each parabola start to mix up
as shown in Fig.~\ref{fig:Maj}(b).

Despite their apparent complexity, we can
analyze these complex spectra in terms of charge and
anyonic excitations. We first establish the recipes to
construct the expected (electric) charge and anyonic spectra from simple considerations. 
Subsequently, we will show that
the exact numerical spectra of the anyonic $t$-$J$ chains can indeed be seen as a subtle sum of the
above two.

The Bethe-Ansatz results~\cite{ogata-shiba,parola-sorella-92} for the $J\rightarrow 0$
{\it electronic} $t$-$J$ chain suggest that
the anyonic excitation spectrum $E_{\rm anyon} (m)$ of the itinerant chain
should be that of $N$ anyons localized on a {\it squeezed} periodic chain 
of length $N=\rho L$, where integers $m$ label the eigenstates
of momenta $k_m=2\pi i_m/N$ for $i_m\in\mathbb{Z}$.
Such a spectrum can be computed by exact diagonalization
and agrees very well with the CFT predictions, even on small chains ($N=16$).
In particular, it shows a linear zero energy mode at $K=0$ and at a characteristic momentum
$k_a=\pi$ (for $J>0$).
The velocity of this spectrum is renormalized by a factor $\gamma$, which is  $\gamma\simeq 0.5$ for $\rho=2/3$ and sufficiently small $J$.

To construct the expected charge excitation spectrum at finite $J$,
we use our understanding of the $J = 0$ limit.
Gradually turning on $J$,
one can adiabatically follow the original parent states at momenta $K_p$
(still labeled for $J\ne 0$ by the same set of integers $p$).
As for $J=0$, changing the momentum of a charge excitation by $\delta K$ can be done by introducing
an external flux $\phi_{\rm ext}=\delta K/\rho$. Hence, by introducing
{\it twisted boundary conditions},
one can {\it compute}
the new (almost parabolic) branch of excitations ${\tilde E}_{\rm charge}(p,\phi_{\rm ext})$
(renormalized with respect to Eq.~(\ref{Eq:charge0})) associated to each parent excitation.

Inspired by the rules for adding holon and spinon spectra in the $J\rightarrow 0$ electronic Bethe-Ansatz,
we now propose to construct the full theoretical excitation spectrum by adding 
the two above spectra as
\begin{equation}
E_{p,m} = {\tilde E}_{\rm charge} (p, k_m) + E_{\rm anyon} (m)
\label{eq:sp-ch-spec}
\end{equation}
and by adding momenta as $K_{p,m}=K_p+\rho k_m$.
Importantly, the phase shift experienced by the (charged) ``holons" coincides
with the total momentum $k_m$
of the anyonic eigenstates defined on the squeezed (undoped) chain.
This generates a subtle coupling between charge and anyonic excitations
that we shall discuss later on.

We now wish to verify that
the proper assignments of the true energy levels according to the form
given by Eq.~(\ref{eq:sp-ch-spec}) can indeed be made accurately. One proceeds by sequentially constructing
the sets of levels that correspond to increasing the charge index $p$. The zero-momentum GS of the model ($p=0$)
leads to the first charge branch (with $E_{\rm anyon}=0$)
when adding a flux $\phi_{\rm ext}$ to the system.
Two ``secondary" parent charge excitations corresponding to exact eigenstates ($\phi_{\rm ext}=0$) of the system
with momenta $K_p=\pm K_c$ ($p=1,2$) lie on
the {\it same} $p=0$ branch at $\phi_{\rm ext}=\pm 2\pi$ as seen in Fig.~\ref{fig:Maj}(b).
It is then possible to construct the expected set of combined
charge plus anyonic excitations $E_{0,m}$ adjusting
the renormalization factor $\gamma$ to get the best fit to a subset of the exact energy levels.
Although there is only one free parameter, we obtain excellent agreement
between the two sets of levels, especially in the small $J$ limit
shown in Fig.~\ref{fig:Maj}(a) where {\it all} anyonic excitations can be assigned very accurately
up to energies of several $J$.
Also for larger $J$, 
low-energy combined charge-anyon excitations $E_{p,m}$ can be identified up to $p=11$ 
as shown in Fig.~\ref{fig:Maj}(b).

{\it Conclusion and outlook --}
Our model  of itinerant anyons on one-dimensional chains can be generalized to other types of anyons (such as Fibonacci anyons or SU(2)$_k$ anyons, and even spin-$1/2$ fermions), and more than just the lowest energy quasiparticles can be included \cite{long-2012}. Our conclusions apply more generally to all these examples.
In particular, we have shown that, in one dimensional chains itinerant non-Abelian anyons fractionalize into
charge and anyonic degrees of freedom, generalizing the spin-charge separation effect of electronic Luttinger liquids.

However, on a periodic ring, the (charged) ``holons" experience a phase shift that coincides
with the total momentum of the anyonic eigenstates.
This generates a subtle topological coupling between charge and anyonic excitations at the microscopic level.
Since the energy shift induced by the twisted boundary condition is of order $D /L$ for the lowest excitations, where $D$ is the Drude weight, this coupling vanishes in a macroscopic view of the low energy limit of an infinite ring, but remains significant at energies of order $J$.
The same phenomenon also appears 
in electronic systems. It has, however, not yet been studied for such systems in detail, presumably because of strong finite size corrections in the spectra due to marginal operators. This provides another example of how studying generalized anyonic models can also contribute to our understanding of electronic models.

In the context of non-Abelian quantum Hall liquids, the emergent separation of charge and anyonic degrees of freedom {\it a posteriori} justifies the treatment of quantum Hall edge modes by product theories like ${\rm Ising} \times \text{U}(1)_4$, with different velocities for the charge and
neutral sector, despite the intrinsic coupling of electric charge $e/4$ and topological charge
$\sigma$ in the fundamental $(\sigma,e/4)$ quasiparticle.

An interesting feature of the electronic $t$-$J$ model is its supersymmetric point which
can be solved by Bethe-Ansatz \cite{w88,bb90}. It will be interesting to investigate
if the anyonic models exhibit an analogous supersymmetry.

We thank A. Feiguin, A.W.W. Ludwig, C. Nayak, and S. Trebst for useful discussions. We acknowledge the hospitality and support of the Aspen Center for Physics under grant number NSF 1066293 and the Pauli Center at ETH Zurich. DP acknowledges support by the French Research Council (Agence Nationale de la Recherche) under grant No. ANR 2010 BLANC 0406-01 and is grateful to Nicolas Renon at CALMIP (Toulouse, France) 
for support in the use of the Altix SGI supercomputer.



\begin{thebibliography}{99}

\bibitem{anderson} P. W. Anderson, Phys. Rev. {\bf 164}, 352 (1967).

\bibitem{t50}
S.~Tomonaga,
Prog. Theor. Phys. {\bf 5}, 544 (1950).

\bibitem{l63}
J. M.~Luttinger,
J. Math. Phys. {\bf 15}, 609 (1963).

\bibitem{h81}
F. D. M.~Haldane,
J. Phys. C {\bf 14}, 2585 (1981).

\bibitem{wen-91} X.~G.~Wen, Int. J. Mod. Phys. B {\bf 5}, 1641 (1991).

\bibitem{Leinaas77} J. M. Leinaas and J. Myrheim, Nuovo Cimento B {\bf 37}, 1 (1977).

\bibitem{Laughlin83} R. B.~Laughlin, Phys. Rev. Lett. {\bf 50}, 1395 (1983).
%
\bibitem{arovas-84prl722} D. Arovas, J.~R.~Schrieffer, and F.~Wilczek, Phys. Rev. Lett. {\bf 53},
722 (1984).
%
\bibitem{bonderson-2011aa}
P. Bonderson, V. Gurarie and C. Nayak, Phys. Rev. B
{\bf 83}, 075303 (2011); 
see also e.g. C.~Nayak and F.~Wilczek, Nucl. Phys. B {\bf 479}, 529 (1996);
V.~Gurarie, C.~Nayak, Nucl. Phys. B {\bf 506}, 685 (1997);
N.~Read, Phys. Rev. B {\bf 79}, 045308 (2009).
%

\bibitem{non-abelian-anyons} G. A. Goldin, R. Menikoff, and D.~H.~Sharp, Phys. Rev. Lett.
{\bf 54}, 603 (1985);
K.~Fredenhagen, K.~H.~Rehren, and B.~Schroer, Commun.
Math. Phys. {\bf 125}, 201 (1989);
J.~Fr\"ohlich and F.~Gabbiani, Rev. Math. Phys. {\bf 2}, 251 (1990).

\bibitem{moore-read}
G. Moore and N. Read, Nucl. Phys. B {\bf 360}, 362 (1991).

\bibitem{read-rezayi} N. Read and E.~Rezayi, Phys.Rev. B {\bf 59} 8084 (1999).

\bibitem{Lee07} S.-S. Lee, S.~Ryu, C.~Nayak and M.~P.~A.~Fisher,
Phys. Rev. Lett. {\bf 99}, 236807 (2007). 
%

\bibitem{Levin07} M.~Levin, B.~I.~Halperin and B.~Rosenow,
Phys. Rev. Lett. {\bf 99}, 236806 (2007).

\bibitem{Bonderson07d} P. Bonderson and J.~K.~Slingerland, Phys. Rev. B
{\bf 78}, 067836 (2008). 

\bibitem{majoranas} 
N. Read and D. Green, Phys. Rev. B {\bf 61}, 10267 (2000);
A.~Y.~Kitaev, Phys.-Usp. {\bf 44}, 131 (2001);
L.~Fu and C.~L.~Kane, Phys. Rev. Lett. {\bf 100}, 096407 (2008).
Here, Ising anyons are often referred to as Majorana bound states. 

\bibitem{kitaev-2003}
A. Kitaev, Ann. Phys. {\bf 2}, 303 (2003).

\bibitem{ZhangRice} F.-C. Zhang and T. M. Rice, Phys. Rev. B {\bf 37}, 3759 (1988).

\bibitem{Hubbard_model} J. Hubbard, Proc. Roy. Soc. London A {\bf 276}, 238 (1963); M.~C.~Gutzwiller, 
Phys. Rev. Lett {\bf 10}, 159 (1963); J.~Kanamori, Prog. of Theor. Phys. (Kyoto) {\bf 30}, 275 (1963).

\bibitem{ogata-shiba} M.~Ogata and H.~Shiba, Phys. Rev. B {\bf 41}, 2326Ð2338 (1990);
H.~Shiba and M.~Ogata, Int. J. of Mod. Phys. B {\bf 5}, 31 (1991).

\bibitem{maekawa} C.~Kim et al., Phys. Rev. Lett. {\bf 77}, 4054 (1996).

\bibitem{note1} 
Nevertheless, the neutral excitation in the $\nu=5/2$ FQHS appears to be a stable excitation, lower 
in energy that the $e/4$ quasihole-quasielectron pair;
P. Bonderson, A. E. Feiguin, C. Nayak, Phys. Rev. Lett. {\bf 106}, 186802 (2011).

\bibitem{ftl07}
A.~Feiguin {\em et al.},
Phys. Rev. Lett. {\bf 98}, 160409 (2007).

\bibitem{Bonderson09b} P. Bonderson, Phys. Rev. Lett. {\bf 103}, 110403 (2009). 
%

\bibitem{difrancesco} P. Di Francesco, P. Mathieu, and D. S\'en\'echal, {\it Conformal Field
Theory} (Springer-Verlag New York, Inc., 1997).

\bibitem{parola-sorella-92} A. Parola and S. Sorella
Phys. Rev. B {\bf 45}, 13156 (1992).

\bibitem{note2}
For the $\nu=1/2$ Moore-Read state, there is a total Abelian phase of $\phi_{\rm ext}=\pi/4$, combining the magnetic flux and an Abelian twist factor of the anyons.

\bibitem{long-2012} D. Poilblanc, M. Troyer, E. Ardonne, A. E. Feiguin and P. Bonderson,
in preparation.

\bibitem{w88}
P. B.~Wiegmann,
Phys. Rev. Lett. {\bf 60}, 821 (1988).

\bibitem{bb90}
P. A.~Bares, G.~Blatter,
Phys. Rev. Lett. {\bf 64}, 2567 (1990).

\end{thebibliography}
\end{document}